\documentclass[12pt]{revtex4-1}
\usepackage{graphicx}
\usepackage[usenames]{color}
\usepackage{color}
\usepackage[sort&compress]{natbib}
\usepackage{amsmath}
\usepackage{amssymb}
\allowdisplaybreaks[1]

\begin{document}

\title{Reaction Kinetics and Mechanism of Magnetic Field Effects in Cryptochrome}

\author{Ilia A. Solov'yov$^{1}$\footnote{E-mail: ilia@illinois.edu}}
\author{Klaus Schulten$^{1,2}$\footnote{E-mail: kschulte@ks.uiuc.edu}}%

\address{$^{1}$Beckman Institute for Advanced Science and Technology, University of Illinois at Urbana-Champaign, 405 N. Matthews Ave., 61801 Urbana, Illinois, USA}
\address{$^{2}$Department of Physics, University of Illinois at Urbana-Champaign, Urbana, Illinois, USA}

\begin{abstract}
Creatures as varied as mammals, fish, insects, reptiles, and birds have an intriguing `sixth' sense that allows them to orient themselves in the Earth's magnetic field. Despite decades of study, the physical basis of this magnetic sense remains elusive. A likely mechanism is furnished by magnetically sensitive radical pair reactions occurring in the retina, the light-sensitive part of animal eyes. A photoreceptor, cryptochrome, has been suggested to endow birds with magnetoreceptive abilities as the protein has been shown to exhibit the biophysical properties required for an animal magnetoreceptor to operate properly. Here, we propose a theoretical analysis method for identifying cryptochrome's signaling reactions involving comparison of measured and calculated reaction kinetics in cryptochrome. Application of the method yields an exemplary light-driven reaction cycle, supported through transient absorption and electron-spin-resonance observations together with known facts on avian magnetoreception. The reaction cycle permits one to predict magnetic field effects on cryptochrome activation and deactivation. The suggested analysis method gives insight into structural and dynamic design features required for optimal detection of the geomagnetic field by cryptochrome and suggests further experimental and theoretical studies.
\end{abstract}

\maketitle

\section{Introduction}

Migratory birds travel annually thousands of kilometers, navigating by using various cues, including the Earth's magnetic field \cite{MOUR2005,WILT2006A,IliaKlausWalter,Wiltschko72,Merkel65}; non-migrant bird species utilize the geomagnetic field similarly to find their way back to the breeding nest \cite{Freire2008,Voss2007,Keeton1971,Wiltschko81,Wiltschko98}. But how do birds detect the geomagnetic field? This question focuses on one of the most fascinating unsolved mysteries of sensory biology. At a first glance, tiny iron-oxide particles detected in the upper beak of some bird species \cite{Fleissner07a,Ilia07a,Ilia08a,Ilia09b,Kirschvink01,Falkenberg2010} provide a natural explanation for the avian magnetic compass sense, but behavioral studies revealed that the ability of night-migrating birds to perform magnetic compass orientation is affected by the ambient light \cite{WILT1993,WILT2001,Wiltschko2004,Muheim2002,Wiltschko2008}, which does not penetrate through the beak skin. The latter observation lead to the suggestion that a photochemical reaction in the bird's eyes produces spin-correlated radical pairs that act as the sensor embodying a magnetic field-dependent signaling cascade \cite{SCHU78C,RITZ2000,SOLO2007,Ilia09,SOLO2010,Rodgers2009}. This suggestion is based on the observation that the recombination reactions of spin-correlated radical pairs, can be magnetic field-sensitive  \cite{SalikhovBook2,Steiner89}. The radical pair hypothesis gained stronger support when it was experimentally demonstrated that the magnetic compass in birds is still functioning properly after the iron-mineral-based receptors in the beak are deactivated \cite{Beason1996,Wiltschko2009B,Wiltschko1994,Zapka2009}.

Spin-correlated radical pairs are typically created in a photochemical reaction that produces radicals from a molecular precursor in either an electronic singlet (S) or triplet (T) state. Under the influence of intramolecular electron-nuclear hyperfine interactions, the radical pair oscillates between the S and T states, a process known as $\mathrm{S}\leftrightarrow\mathrm{T}$ interconversion. External static magnetic fields affect the rate of this process and, hence, alter the yields of the respective reaction products formed from the S and T radical pairs \cite{SCHU78A,SCHU78D,SCHU86C,SCHU76C,SCHU84,WERN78,Steiner89}. During the last years the radical-pair hypothesis of the avian magnetic compass gained significant experimental \cite{THAL2005,Keary2009,RITZ2004,Ritz2009,Wiltschko2007} and theoretical \cite{Rodgers2009,SCHU78C,RITZ2000,SOLO2007,Ilia09,Ilia08b,SOLO2010,CINT2003} support. The first experimental proof-of-principle, demonstrating that under a static magnetic field, as weak as that of the Earth, a chemical reaction can act as a magnetic compass by producing detectable changes in the chemical product yield, was achieved for a carotenoid-porphyrin-fullerene model system \cite{Maeda2008}.

However, for the radical-pair mechanism to play a role in magnetoreception, molecules with certain biophysical characteristics must exist in the eyes of migratory birds. Cryptochromes \cite{AHMA1996,CASH1999,SANC2003,Ahmad1993}, a class of photoreceptor proteins, were proposed as the host molecules for the crucial radical pair co-factors that putatively act as a primary magnetoreceptor \cite{RITZ2000}. Cryptochrome is a signaling protein found in a wide variety of plants and animals \cite{AHMA1996,CASH1999,SANC2003}. Its role varies widely among organisms, from the entrainment of circadian rhythms in vertebrates to the regulation of hypocotyl elongation and anthocyanin production in plants \cite{LIN2005,PART2005,CHRI2001}. It has recently been demonstrated, using UV/visible transient absorption and electron paramagnetic resonance spectroscopy \cite{Liedvogel2007,Biskup2009}, that vertebrate cryptochromes form long-lived radical pairs, involving a flavin radical and a radical derived from a redox-active amino acid. This observation demonstrates that cryptochrome harbors the type of radical pair needed for magnetic compass action and suggests that cryptochromes, through $\mathrm{S}\leftrightarrow\mathrm{T}$ interconversion, are influenced by an external magnetic field. Radical pairs seen in DNA photolyase \cite{Kao2008,Prytkova2007,Henbest2008,BYRD2003,Gindt1999,AUBE1999,AUBE2000}, a light sensitive DNA repair enzyme closely related to cryptochrome, support the observation \cite{Liedvogel2007,Biskup2009} in cryptochrome; indeed a recent study \cite{Henbest2008} demonstrated a magnetic field effect in DNA photolyase, suggesting a similar effect to exist in cryptochrome.

It has been verified that cryptochromes exist in the eyes of migratory birds \cite{MOUR2004,MOLL2004,Liedvogel2007,Liedvogel2010,Niessner2011}, and that at least some cryptochrome-containing cells within the retina are active at night when the birds perform magnetic orientation in the laboratory \cite{MOUR2004,Niessner2011}. Furthermore, a distinct part of the forebrain, which primarily processes input from the eyes, is highly active at night in night-migratory garden warblers (\textit{Sylvia borin}) and European robins (\textit{Erithacus rubecula}) \cite{MOUR2005b,Liedvogel07,Feenders2008,Zapka2009,Heyers2007}. In summary, many findings are consistent with the hypothesis that magnetic compass detection in migratory birds takes place in the eye \cite{Zapka2009,Hein2010,WILT2002,Hein2011,Wiltschko2011,Stapput2010} and that cryptochromes are the primary magnetoreceptors.
However, despite the success of these findings, a completely satisfactory description of the mechanism of the magnetic field effect in cryptochrome is still missing.

\begin{figure}[!t]
\begin{center}
\includegraphics[width=11.7cm]{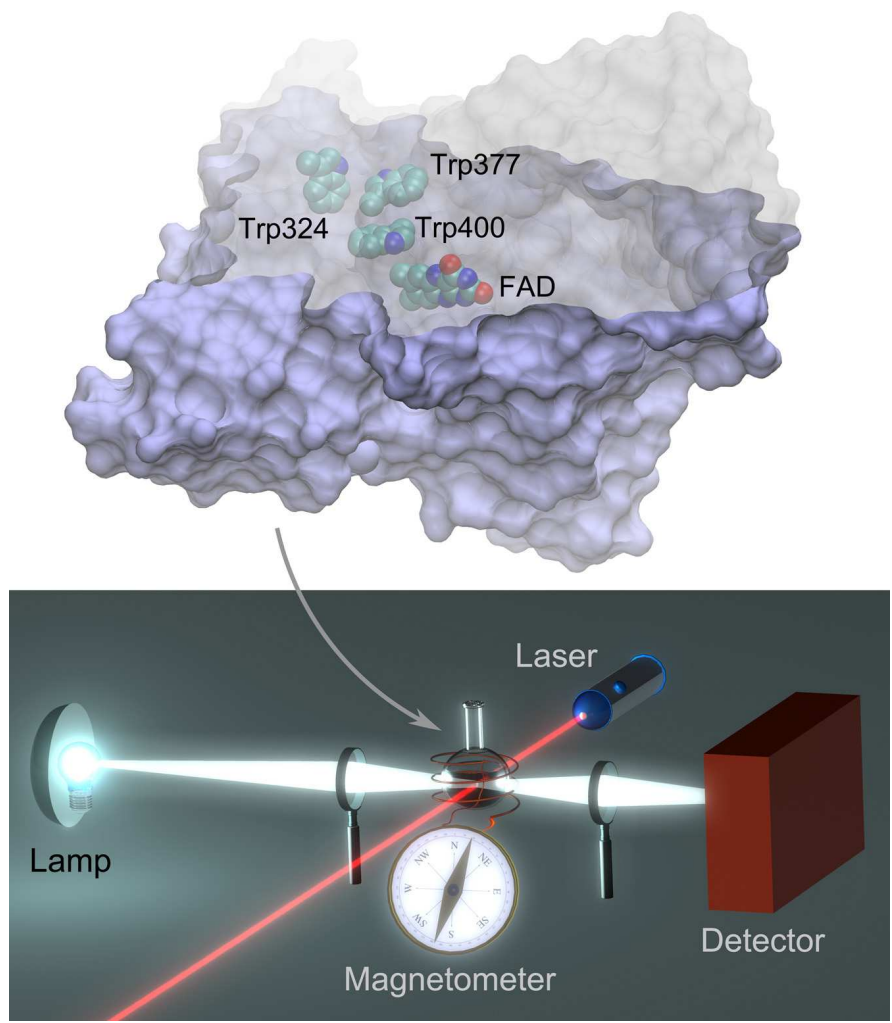}
\end{center}
\caption{
{\bf \textit{Arabidopsis thaliana} cryptochrome in transient absorption experiment.} (Top) The structure of cryptochrome-1 from \textit{Arabidopsis thaliana} \cite{BRAU2004} is shown together with the highlighted flavin cofactor (FAD) and the tryptophan triad Trp400, Trp377, Trp324. (Bottom) Transient absorption of cryptochrome is probed by means of the pump-probe experiment, as described in experiment \cite{Liedvogel2007}. The sample containing cryptochrome is irradiated with a pulsed laser beam (red) that generates a measurable concentration of excited states ($\mathrm{FAD^{*}}$) in the system. $\mathrm{FAD^{*}}$ decays then in a series of intermediates back to FAD, while some of the intermediates being probed by the probe beam (white) used to measure the absorption spectrum of transient species. Additionally, magnetic field effects in cryptochrome can be studied if the sample is subjected to an external magnetic field.}
\label{fig:experiment}
\end{figure}

The process of cryptochrome photoactivation was discussed earlier and several reaction schemes were proposed \cite{Bouly2007,GIOV2003,Kao2008,Hoang2008,Shirdel2008,Liedvogel2010}. However, the proposed schemes are incomplete, usually accounting for only some of several observations.
Cryptochrome binds internally the chromophore flavin adenine dinucleotide (FAD) \cite{SANC2003,BRAU2004,Shirdel2008,Hoang2008,Liedvogel2010,Bouly2007,Pokorny2008}. At least in plant cryptochromes from \textit{Arabidopsis thaliana} \cite{GIOV2003,Liedvogel2007,Bouly2007,Hoang2008,Banerjee2007,Langenbacher2009} blue light leads to conversion of the fully oxidized FAD to the semireduced FADH$^{\bullet}$ form, the latter representing the signaling state. The conversion happens in the course of a light-induced electron transfer reaction involving FAD and a chain of three tryptophan amino acids that bridge the space between FAD and the protein surface \cite{ZEUG2005,Liedvogel2007,Biskup2009}, as illustrated in Fig.~\ref{fig:experiment}. Atomic level structures of plant cryptochrome are known for \textit{Arabidopsis thaliana} cryptochrome-1 \cite{BRAU2004} and \textit{Arabidopsis thaliana} DASH-type cryptochrome-3 \cite{Pokorny2008}.
In cryptochromes from insects, light excitation leads to formation of a flavin anion radical, FAD$^{\bullet-}$ \cite{Langenbacher2009,Berndt2007}. It is currently under debate whether the anion radical represents the signaling state or whether it is only a functionally insignificant short-lived intermediate \cite{Berndt2007,Shirdel2008,Oeztuerk2008}. The difference in the photocycles of plant and insect cryptochromes illustrates that detailed analysis of the transient states in cryptochromes is needed in light of the fact that major variations arise between different cryptochromes, even regarding the nature of their signaling states. Such analysis can only be performed \textit{in vitro} on cryptochromes extracted from specific organisms.

Unfortunately, relative to plant and insect cryptochromes, little is known at present about structure and photocycle of avian cryptochromes. Therefore, we study here a model ``cryptochrome'' from a general perspective. Our goal is to demonstrate what observations are needed for resolving photoactivation and magnetoreception of the protein. By suggesting an analysis method that deduces from experimental observations a reaction scheme, it is our hope that further physico-chemical experiments will be initiated on cryptochrome photoactivation that ultimately pinpoint the complete photoreaction and signaling process in cryptochrome and, in particular, establish when and why the anionic, FAD$^{\bullet-}$, and the neutral, FADH$^{\bullet}$, radicals are formed.

\begin{figure}[!tp]
\begin{center}
\includegraphics[width=12.cm]{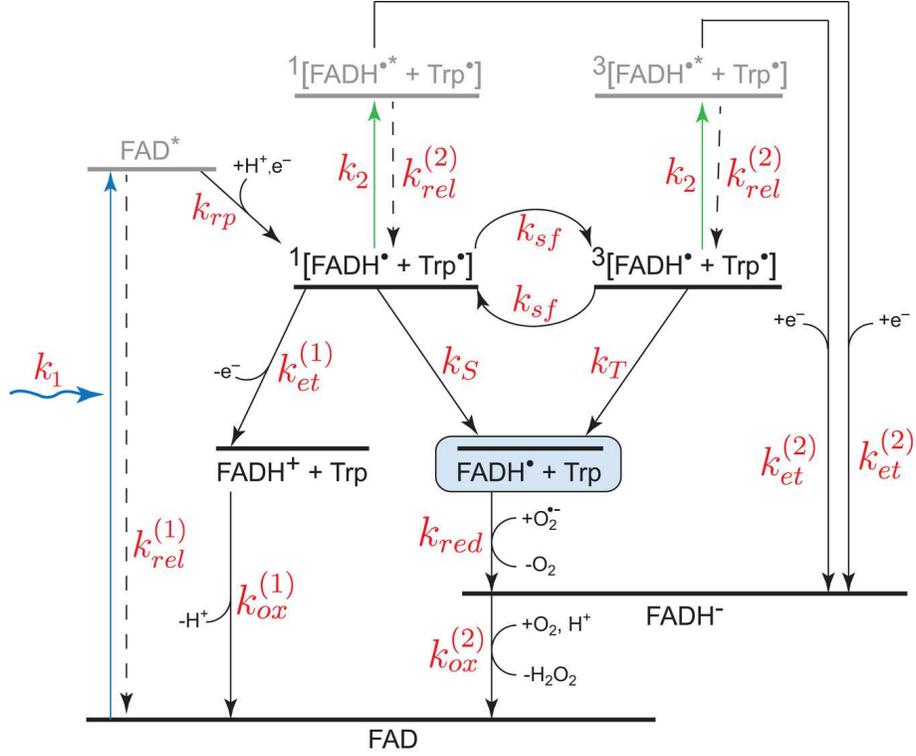}
\end{center}
\caption{
{\bf Cryptochrome activation and inactivation reactions.} Cryptochrome is activated through absorbing a blue-light photon by the flavin cofactor, responsible for protein's signaling. Initially, the flavin cofactor in cryptochrome is present in its fully oxidized FAD state. After absorbing a photon, FAD becomes promoted to an excited $\mathrm{FAD}^{*}$ state. $\mathrm{FAD}^{*}$ is then protonated and receives an electron from a nearby tryptophan (see Fig.~\ref{fig:experiment}), leading to the formation of the $[\mathrm{FADH}^{\bullet} + \mathrm{Trp}^{\bullet}]$ radical pair, which exists in singlet and the triplet overall electron spin states, denoted as $^1[\; \cdots \;]$ and $^3[\; \cdots \;]$, respectively. The $\mathrm{Trp^{\bullet}}$ radical may receive an additional electron from a nearby tyrosine \cite{Liedvogel2007,GIOV2003}, or become deprotonated \cite{SOLO2007,BYRD2004}, quenching the radical pair and fixing the electron on the $\mathrm{FADH}^{\bullet}$ cofactor. Under aerobic conditions, $\mathrm{FADH^{\bullet}}$ slowly reverts back to the initial inactive FAD state through the also inactive $\mathrm{FADH^{-}}$ state of the flavin cofactor. Before the $[\mathrm{FADH}^{\bullet} + \mathrm{Trp}^{\bullet}]$ radical pair in cryptochrome is quenched, the electron of the $\mathrm{FADH^{\bullet}}$ radical can back-transfer to the $\mathrm{Trp}^{\bullet}$ radical, thereby also ending the signaling state of the protein. The electron back-transfer leads to the formation of $\mathrm{FADH^{+}}$ and can only occur if the spins of the two unpaired electrons  in the radical pair $[\mathrm{FADH}^{\bullet} + \mathrm{Trp}^{\bullet}]$ are in an overall singlet state. While the flavin cofactor is in its $\mathrm{FADH^{\bullet}}$ state, cryptochrome may absorb a second blue-green photon, thereby transferring it into a non-signaling state with the flavin cofactor in the $\mathrm{FADH^{-}}$ conformation.}
\label{fig:reaction}
\end{figure}

One can safely state that, in spite of numerous experimental observations of intermediate states in cryptochrome, the actual photoreaction is still not fully understood \cite{Biskup2009,GIOV2003,Phillips2010,Bouly2007,Kao2008,Liedvogel2010,Hoang2008,Liedvogel2007,Liu2010,Langenbacher2009,Brazard2010,Chaves2011}. Figure~\ref{fig:reaction} shows a scheme of cryptochrome activation and inactivation. The scheme incorporates the key observation that the flavin cofactor in cryptochrome is observed in three interconvertible redox forms, $\mathrm{FAD}$, $\mathrm{FADH}^{\bullet}$, and $\mathrm{FADH}^{-}$ \cite{Bouly2007,ODay08,Kao2008,Banerjee2007,Biskup2009,Liu2010}. In this scheme the FAD form is inactive (non-signaling) and accumulates to a high level in the dark \cite{Phillips2010,Hoang2008,GIOV2003,Shirdel2008,Bouly2007}. Blue light triggers photoreduction of FAD to establish a photoequilibrium that favors $\mathrm{FADH}^{\bullet}$ over $\mathrm{FAD}$ or $\mathrm{FADH}^{-}$ \cite{Phillips2010,Hoang2008,GIOV2003,Shirdel2008,Bouly2007}. As pointed out above, in plant cryptochromes, signaling is linked to formation of the $\mathrm{FADH}^{\bullet}$ state. This state can absorb a second, blue-green light photon, in which case $\mathrm{FADH}^{\bullet}$ is converted to the fully reduced, inactive form $\mathrm{FADH}^{-}$: the latter reoxidizes in the dark to the original FAD resting state \cite{Bouly2007,ODay08,Kao2008,Henbest2008,Ilia09,Liu2010}.

The tryptophan triad in cryptochrome, depicted in Fig.~\ref{fig:experiment}, is crucial for the functioning of the protein as a primary electron donor \cite{Liedvogel2007,Biskup2009,GIOV2003}. In \textit{Arabidopsis thaliana} the triad consists of Trp324, Trp377, and Trp400 as shown in Fig.~\ref{fig:experiment}. In cryptochromes from garden warbler (\textit{Sylvia borin}) \cite{Liedvogel2007}, \textit{Drosophila melanogaster} \cite{Biskup2009,Hoang2008,Shirdel2008} and \textit{Homo sapiens} \cite{Biskup2009,Hoang2008} the tryptophan-triad is conserved \cite{SOLO2007,Liedvogel2007,Kao2008,Biskup2009}. Before light activation, the flavin cofactor of cryptochrome is present in its fully oxidized FAD state (see Fig.~\ref{fig:reaction}). FAD absorbs blue light, being thereby promoted to an excited state, $\mathrm{FAD}^{*}$, which is then protonated, likely from a nearby aspartic acid \cite{KOTT2006}, and receives an electron from the nearby tryptophan (see Fig.~\ref{fig:experiment}), leading to the formation of a $[\mathrm{FADH}^{\bullet} + \mathrm{Trp400}^{\bullet}]$ radical pair \cite{SOLO2007}. This radical pair is further transformed into a $[\mathrm{FADH}^{\bullet} + \mathrm{Trp324}^{\bullet}]$ radical pair (denoted in Fig.~\ref{fig:reaction} as $^{1,3}[\mathrm{FADH}^{\bullet} + \mathrm{Trp}^{\bullet}]$) via sequential electron transfer, also called paramagnetic-diamagnetic exchange \cite{SCHU85B}, involving the tryptophan triad chain \cite{SOLO2007,Biskup2009,Kao2008}. The $[\mathrm{FADH}^{\bullet} + \mathrm{Trp}^{\bullet}]$ radical pair exists in singlet and triplet states, denoted in Fig.~\ref{fig:reaction} as $^{1}[\mathrm{FADH}^{\bullet} + \mathrm{Trp}^{\bullet}]$ and $^{3}[\mathrm{FADH}^{\bullet} + \mathrm{Trp}^{\bullet}]$, respectively.

The $[\mathrm{FADH}^{\bullet} + \mathrm{Trp}^{\bullet}]$ radical pair is associated with the cryptochrome signaling state, since it involves the flavin cofactor in the $\mathrm{FADH}^{\bullet}$ redox state. We note, though, that the semiquinone $\mathrm{FADH}^{\bullet}$ state of FAD was identified as the signaling state in plant cryptochrome from \textit{Arabidopsis thaliana} \cite{Bouly2007,Banerjee2007,Langenbacher2009}, but that the signaling state of avian cryptochrome is still unknown. In our study we generalize the observation in \textit{Arabidopsis thaliana} to birds; if future studies will demonstrate that a different redox state of the flavin cofactor is governing the signaling behavior of avian cryptochromes, the suggested scheme can be readily adapted. We also note that the $\mathrm{Trp^{\bullet}}$ radical may receive an additional electron from a nearby tyrosine \cite{Liedvogel2007,GIOV2003}, or become deprotonated \cite{SOLO2007,BYRD2004}, stabilizing the $\mathrm{FADH}^{\bullet}$ redox state. The charge state of the tryptophan radical in the radical pair state might be crucial for the energetics and kinetics of the reaction pathway, but it is not essential in a methodological sense, the main focus of the present study. The lifetime of the radical pair state $[\mathrm{FADH}^{\bullet} + \mathrm{Trp}^{\bullet}]$ in cryptochrome is $\gtrsim6\ \mathrm{\mu s}$, according to transient EPR measurement \cite{Biskup2009}.

Under aerobic conditions, the $\mathrm{FADH^{\bullet}}$ redox state reverts back to the initial FAD state \cite{GIOV2003,Bouly2007,Liedvogel2007,ODay08} (see Fig.~\ref{fig:reaction}). This process is not well understood, but seems to occur on a millisecond time scale \cite{GIOV2003,Bouly2007,Liedvogel2007}. It has been suggested that the back-reaction involves the superoxide radical O$_{2}^{\bullet-}$ \cite{Ritz2009}, the reaction evolving through the inactive $\mathrm{FADH^{-}}$ state of the flavin cofactor \cite{Ilia09,Ritz2009,Hore2009}, as depicted in Fig.~\ref{fig:reaction}. The lifetime of the $\mathrm{FADH^{-}}$ state, populated such, should be long enough to allow its detection in transient absorption measurements \cite{Biskup2009,Liedvogel2007}.

Quenching of the $[\mathrm{FADH}^{\bullet} + \mathrm{Trp}^{\bullet}]$ radical pair in cryptochrome can also arise through electron back-transfer from $\mathrm{FADH^{\bullet}}$ to a tryptophan. This back-transfer, leading to formation of $\mathrm{FADH^{+}}$ (see Fig.~\ref{fig:reaction}), can only occur if the spins of the two unpaired electrons are in an overall singlet state. An external magnetic field can influence the overall electron spin state through the Zeeman interaction acting jointly with hyperfine coupling with hydrogen and nitrogen atoms \cite{SCHU76C,SOLO2007,Ilia09,Ilia08b}. If the overall electron spin state of $[\mathrm{FADH}^{\bullet} + \mathrm{Trp}^{\bullet}]$ is triplet, electron back-transfer and formation of $\mathrm{FADH^{+}}$ cannot occur, extending the time cryptochrome stays in its signaling state. The $\mathrm{FADH^{+}}$ state has a short lifetime and decays quickly to the fully oxidized FAD configuration via deprotonation.

According to recent experimental measurements \cite{Hoang2008,Bouly2007,Phillips2010,Kao2008} and as pointed out above, cryptochrome may absorb an additional blue-green light photon, thereby transferring itself into a non-signaling state, namely with the flavin cofactor in the $\mathrm{FADH^{-}}$ redox state. The process involves the transition $\mathrm{FADH^{\bullet}}\rightarrow\mathrm{FADH^{\bullet*}}\rightarrow\mathrm{FADH^{-}}$, depicted in Fig.~\ref{fig:reaction}. After excitation, the flavin cofactor in the $\mathrm{FADH^{\bullet*}}$ conformation  has an electron vacancy, allowing it to accept an electron from a nearby tryptophan and, thereby, be transformed into $\mathrm{FADH^{-}}$. In order to stabilize $\mathrm{FADH^{-}}$, the terminating tryptophan of the triad should be in its radical state. In the absence of such tryptophan radical prior to $\mathrm{FADH^{\bullet}}$ excitation, the excited electron in $\mathrm{FADH^{\bullet*}}$ will back-transfer to the electron vacancy in tryptophan as soon as $\mathrm{FADH^{-}}$ is formed.

If cryptochrome is indeed the avian magnetoreceptor protein, its signaling state is expected to be sensitive to a magnetic field, the latter controlling the lifetime of the $\mathrm{FADH^{\bullet}}$ state. In earlier studies we demonstrated that weak magnetic fields, comparable with the geomagnetic field, can alter the lifetime of $\mathrm{FADH^{\bullet}}$ by 5\%-10\% \cite{SOLO2007,Ilia09}. In these earlier studies we considered reactions in isolation, while in the present study we investigate the complete activation cycle. In the following we suggest a light-driven reaction cycle in cryptochrome that exhibits a representative magnetic field effect on the signaling state of the protein. The reaction cycle is supported by transient absorption and electron-spin-resonance observations \cite{Biskup2009,Liedvogel2007} and incorporates known attributes of avian magnetoreception. We demonstrate how the proposed reaction cycle can be compared to experiment.



\section{Methods}

Cryptochrome is a sensory protein \cite{Hoang2008,Liedvogel2010,Bouly2007,Biskup2009,Phillips2010,Liu2010}. Its relevant properties, namely, light activation and deactivation, arise through a complex reaction scheme linking many intermediate states. In case of magnetoreception, cryptochrome activation and deactivation is apparently magnetic field dependent. The long-range goal of our study is to establish cryptochrome's photo-reaction scheme unequivocally, combining measurement and theoretical analysis.

Key for the stated goal are measurements of cryptochrome transient photoabsorption \cite{Liedvogel2007,Biskup2009} that  presently suggest different reaction schemes, consistent with the same experimental observations. For the sake of concreteness we single out one reaction scheme, namely the one shown in Fig.~\ref{fig:reaction}. We will demonstrate how this scheme can be reconciled with observed time-dependent absorption spectra. The method can be applied to variant schemes, too.

The efficiency of light absorption at wavelength $\lambda$ by an absorbing medium is characterized through the absorbance $A(\lambda)$ defined as \cite{SchmidtBook}

\noindent
\begin{equation}
A(\lambda)=d\sum_{i=1}^{N}\varepsilon_i(\lambda)c_i,
\label{eq:BeerLambertLaw}
\end{equation}

\noindent
where $N$ is the number of different light-absorbing components in the system with the concentrations $c_i$ (expressed in $\mathrm{mol\cdot L^{-1}}$) and the molar absorption (extinction) coefficients $\varepsilon_i(\lambda)$ (in $\mathrm{L\cdot mol^{-1}cm^{-1}}$); $d$ is the thickness of the absorbing medium (in $\mathrm{cm}$). The molar absorption coefficients in Eq.~(\ref{eq:BeerLambertLaw}) are wavelength-dependent, resulting in a wavelength-dependence of the sample absorbance. The time evolution of the intermediate states concentrations can be calculated from the set of coupled kinetic equations

\noindent
\begin{align}
\label{eq:rate_equations1}
\frac{\mathrm{d}[\mathrm{FAD}]}{\mathrm{d}t}=&\,-k_{1}[\mathrm{FAD}]+k_{rel}^{(1)}[\mathrm{FAD^{*}}]+k_{ox}^{(1)}[\mathrm{FADH^{+}}]+k_{ox}^{(2)}[\mathrm{FADH^{-}}],\\
\label{eq:rate_equations2}
\frac{\mathrm{d}[\mathrm{FAD^{*}}]}{\mathrm{d}t}=&\,k_{1}[\mathrm{FAD}]-k_{rp}[\mathrm{FAD^{*}}]-k_{rel}^{(1)}[\mathrm{FAD^{*}}],\\
\nonumber
\frac{\mathrm{d}[\mathrm{^1(FADH^{\bullet}Trp^{\bullet})}]}{\mathrm{d}t}=&\,k_{rp}[\mathrm{FAD^{*}}]-\left(k_{et}^{(1)}+k_{2}+k_{S}+k_{sf}\right)[\mathrm{^1(FADH^{\bullet}Trp^{\bullet})}]+\\
\label{eq:rate_equations3}
&k_{sf}[\mathrm{^3(FADH^{\bullet}Trp^{\bullet})}]+k_{rel}^{(2)}[\mathrm{^1(FADH^{\bullet*}Trp^{\bullet})}],\\
\nonumber
\frac{\mathrm{d}[\mathrm{^3(FADH^{\bullet}Trp^{\bullet})}]}{\mathrm{d}t}=&\,k_{sf}[\mathrm{^1(FADH^{\bullet}Trp^{\bullet})}]-\left(k_{2}+k_{T}+k_{sf}\right)[\mathrm{^3(FADH^{\bullet}Trp^{\bullet})}]+\\
\label{eq:rate_equations4}
&k_{rel}^{(2)}[\mathrm{^3(FADH^{\bullet*}Trp^{\bullet})}],\\
\label{eq:rate_equations5}
\frac{\mathrm{d}[\mathrm{^1(FADH^{\bullet*}Trp^{\bullet})}]}{\mathrm{d}t}=&\,k_{2}[\mathrm{^1(FADH^{\bullet}Trp^{\bullet})}]-\left(k_{rel}^{(2)}+k_{et}^{(2)}\right)[\mathrm{^1(FADH^{\bullet*}Trp^{\bullet})}],\\
\label{eq:rate_equations5a}
\frac{\mathrm{d}[\mathrm{^3(FADH^{\bullet*}Trp^{\bullet})}]}{\mathrm{d}t}=&\,k_{2}[\mathrm{^3(FADH^{\bullet}Trp^{\bullet})}]-\left(k_{rel}^{(2)}+k_{et}^{(2)}\right)[\mathrm{^3(FADH^{\bullet*}Trp^{\bullet})}],\\
\label{eq:rate_equations6}
\frac{\mathrm{d}[\mathrm{FADH^{+}}]}{\mathrm{d}t}=&\,k_{et}^{(1)}[\mathrm{^1(FADH^{\bullet}Trp^{\bullet})}]-k_{ox}^{(1)}[\mathrm{FADH^{+}}],\\
\label{eq:rate_equations7}
\frac{\mathrm{d}[\mathrm{FADH^{\bullet}}]}{\mathrm{d}t}=&\,k_{S}[\mathrm{^1(FADH^{\bullet}Trp^{\bullet})}]+k_{T}[\mathrm{^3(FADH^{\bullet}Trp^{\bullet})}]-k_{red}[\mathrm{FADH^{\bullet}}],\\
\nonumber
\frac{\mathrm{d}[\mathrm{FADH^{-}}]}{\mathrm{d}t}=&\,k_{red}[\mathrm{FADH^{\bullet}}]+k_{et}^{(2)}\left([\mathrm{^1(FADH^{\bullet*}Trp^{\bullet})}]+[\mathrm{^3(FADH^{\bullet*}Trp^{\bullet})}]\right)-\\
\label{eq:rate_equations8}
&k_{ox}^{(2)}[\mathrm{FADH^{-}}].
\end{align}

\noindent
Here square brackets denote the concentration of transient states  introduced in Fig.~\ref{fig:reaction}; $[\mathrm{FAD}]$,\, $[\mathrm{FAD^{*}}]$,\,\, $[\mathrm{^1(FADH^{\bullet}Trp^{\bullet})}]$,\,\, $[\mathrm{^3(FADH^{\bullet}Trp^{\bullet})}]$,\,\, $[\mathrm{^1(FADH^{\bullet*}Trp^{\bullet})}]$,\,\, $[\mathrm{^3(FADH^{\bullet*}Trp^{\bullet})}]$,\,\, $[\mathrm{FADH^{\bullet*}}]$, $[\mathrm{FADH^{+}}]$, $[\mathrm{FADH^{\bullet}}]$ and $[\mathrm{FADH^{-}}]$ denote the concentration of $\mathrm{FAD}$, $\mathrm{FAD^{*}}$, $\mathrm{^1[FADH^{\bullet}+Trp^{\bullet}]}$, $\mathrm{^3[FADH^{\bullet}+Trp^{\bullet}]}$, $\mathrm{^1[FADH^{\bullet*}+Trp^{\bullet}]}$, $\mathrm{^3[FADH^{\bullet*}+Trp^{\bullet}]}$, $\mathrm{FADH^{\bullet*}}$, $\mathrm{FADH^{+}}$ +$\mathrm{Trp}$, $\mathrm{FADH^{\bullet}+Trp}$ and $\mathrm{FADH^{-}}$ states, respectively. The rate constants $k_{1}$, $k_{2}$, $k_{rp}$, $k_{sf}$, $k_{S}$, $k_{T}$, $k_{rel}^{(1)}$, $k_{rel}^{(2)}$, $k_{ox}^{(1)}$ , $k_{ox}^{(2)}$, $k_{red}$, $k_{et}^{(1)}$ and $k_{et}^{(2)}$ in Eqs.~(\ref{eq:rate_equations1})-(\ref{eq:rate_equations8}) capture the different processes underlying cryptochrome photoactivation and relaxation shown in Fig.~\ref{fig:reaction}, assuming all processes can be described through first order kinetics. The choice of the appropriate rate constants is discussed in Supplementary Material and is summarized in Tab.~\ref{tab:rates}.

\begin{table}[!ht]
\caption{\bf{Rate constants of various kinetic processes in cryptochrome.}}
\begin{tabular}{| l  |l  |l  |l  |l  |}
\hline
Rate    & Physical process & Char. & Rate ($\mathrm{s^{-1}}$) & Source \\
const.&                  & time           &                 &        \\
\hline
$k_{1}$  & FAD excitation by blue light                                & $1.47\ \mathrm{ns}$    & $6.8\times10^{8}$  & estimate, see Supporting Material\\
$k_{2}$  & $\mathrm{FADH}^{\bullet}$ excitation by green light                  & $1.1\ \mathrm{ns}$ & $9\times10^{8}$  & estimate, see Supporting Material\\
$k_{rp}$      & $^1\mathrm{[FADH^{\bullet}+Trp^{\bullet}}$] radical pair &    &  & \\
              & formation & $30\ \mathrm{ps}$   & $3.3\times10^{10}$ & experiment, \citenum{AUBE2000}\\
$k_{sf}$      & Singlet$\leftrightarrow$triplet interconversion               & $1\ \mathrm{\mu s}$ & $10^{6}$  & estimate, \citenum{Rodgers2009}\\
$k_{S}$       & Singlet state decay                                       & $10\ \mathrm{\mu s}$& $10^{5}$  & consistent with \citenum{Biskup2009}\\
$k_{T}$       & Triplet state decay                                       & $10\ \mathrm{\mu s}$& $10^{5}$  & consistent with \citenum{Biskup2009}\\
$k_{rel}^{(1)}$ & $\mathrm{FAD^{*}}$ relaxation to FAD                                 & $80\ \mathrm{ps}$   & $1.25\times10^{10}$ & experiment, \citenum{BYRD2003}\\
$k_{rel}^{(2)}$ & $\mathrm{FADH^{\bullet*}}$ relaxation to $\mathrm{FADH^{\bullet}}$            & $80\ \mathrm{ps}$   & $1.25\times10^{10}$ & experiment, \citenum{BYRD2003}\\
$k_{ox}^{(1)}$  & $\mathrm{FADH^{+}}$ deprotonation                                    & $10\ \mathrm{ps}$   & $10^{11}$ & experiment,\\
& & & & \citenum{Henbest2008,Bouly2007,Kao2008,Hoang2008}\\
$k_{ox}^{(2)}$  & $\mathrm{FADH^{-}}$ oxidation                                    & $4\ \mathrm{ms}$    & $250$     & experiment, \citenum{Liedvogel2007} \\
$k_{red}$     & $\mathrm{FADH^{\bullet}}$ reduction                                & $14\ \mathrm{ms}$   & $70$      & experiment, \citenum{Liedvogel2007}\\
$k_{et}^{(1)}$  & electron transfer $\mathrm{FADH^{\bullet}\rightarrow Trp^{\bullet}}$ & $10\ \mathrm{\mu s}$& $10^{5}$  & experiment, \citenum{Biskup2009}\\
$k_{et}^{(2)}$  & electron transfer $\mathrm{Trp\rightarrow FADH^{*\bullet}}$         & $38\ \mathrm{ps}$   & $2.6\times10^{10}$ & experiment, \citenum{BYRD2003}\\
\hline
\end{tabular}
\begin{flushleft}
Characteristic time scales and rate constants for the kinetic processes in cryptochrome depicted in Fig.~\ref{fig:reaction}.
\end{flushleft}
\label{tab:rates}
 \end{table}

To demonstrate that the reaction scheme in Fig.~\ref{fig:reaction} is consistent with observed cryptochrome transient absorption \cite{Liedvogel2007} we evaluate the time development of $[\mathrm{FAD}]$, $[\mathrm{FAD^{*}}]$, $[\mathrm{^1(FADH^{\bullet}Trp^{\bullet})}]$, $[\mathrm{^3(FADH^{\bullet}Trp^{\bullet})}]$, $[\mathrm{^1(FADH^{\bullet*}Trp^{\bullet})}]$, $[\mathrm{^3(FADH^{\bullet*}Trp^{\bullet})}]$, $[\mathrm{FADH^{\bullet*}}]$, $[\mathrm{FADH^{+}}]$, and $[\mathrm{FADH^{-}}]$. The evaluation can be achieved by numerical integration of the rate equations (\ref{eq:rate_equations1})-(\ref{eq:rate_equations8}). We assume the initial condition in which holds $\left.[\mathrm{FAD}]\right|_{t=0}=c_0$, while all other intermediate states are unpopulated. $c_0$ is the initial ($t=0$) concentration of cryptochrome in the sample before excitation by a laser pulse. According to the experimental study \cite{Liedvogel2007} holds $c_0\approx20\ \mathrm{\mu M}$.

\section{Results}

In the following the time-dependence of the transient absorption in cryptochrome is studied for the reaction scheme shown in Fig.~\ref{fig:reaction}. The calculated transient absorption spectra based on the scheme are then compared to available experimental data. After the postulated reaction has been validated through comparison with experimental observations, the influence of an external magnetic field on the signaling state of the protein is analyzed.

\subsection*{Light intensity dependence of transient spectra}

\begin{figure}[!t]
\begin{center}
\includegraphics[width=16cm]{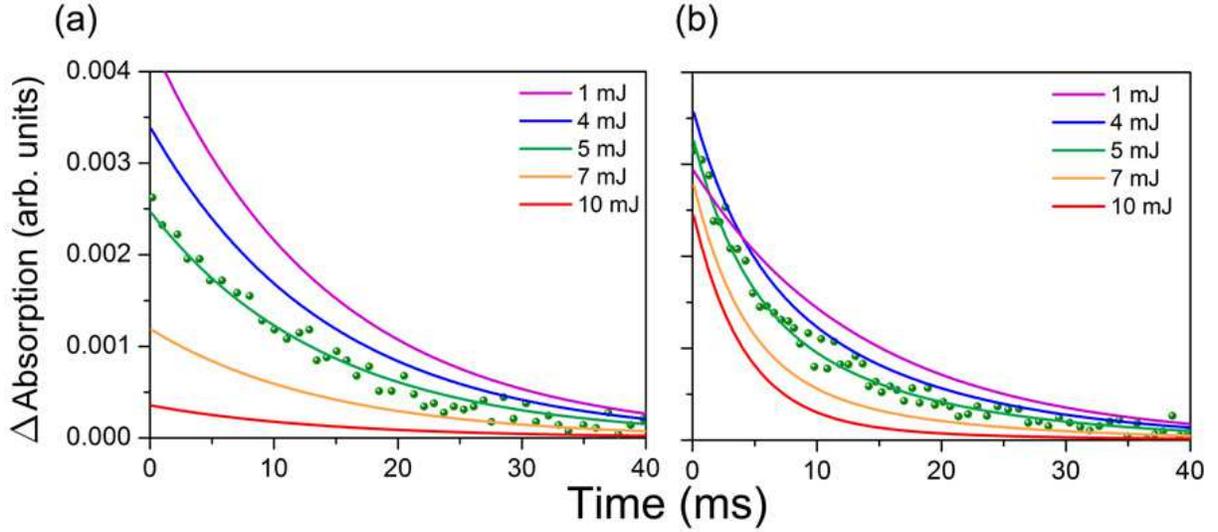}
\end{center}
\caption{
{\bf Cryptochrome transient absorption spectra.} Time profiles of the change in the transient absorption spectra from the absorbance prior laser excitation of the sample calculated for cryptochrome excited by laser pulses of different intensity (see Fig.~\ref{fig:experiment}), for the probe-beam of wavelengths 550-630~nm (plot a), and 490-550~nm (plot b). The laser pulse duration chosen was 5~ns as in experiment \cite{Liedvogel2007}, at 355~nm wavelength. The energies of the pulse used in the calculation are indicated in the inset. The wavelengths used in the calculation were chosen consistent with the experimental measurements of the transient absorption spectra in garden warbler cryptochrome \cite{Liedvogel2007}. Dots correspond to the experimentally measured data recorded for the pulse energy of 5~mJ \cite{Liedvogel2007}.}
\label{fig:transientAbsorptionIntensity}
\end{figure}

To probe the transient states in cryptochrome,  a cryptochrome-containing sample from the retina of garden warbler had been studied \cite{Liedvogel2007}. The transient absorption was measured using the pump-probe experiment depicted in Fig.~\ref{fig:experiment}, where a Nd-YAG laser operating at 355~nm excited the sample, and a 150~W tungsten lamp was used for monitoring reaction intermediates. To study the transient states in cryptochrome during its photoactivation cycle the wavelength of the probe beam was adjusted to allow absorption of either the $\mathrm{FADH}^{-}$ and $\mathrm{FADH}^{\bullet}$ states of the flavin co-factor (absorbing at 490-550~nm), or of the $\mathrm{FADH}^{\bullet}$ state only (absorbing at 550-630~nm). The experimentally recorded time profiles of the transient absorption spectra for the two regimes are shown in Fig.~\ref{fig:transientAbsorptionIntensity}. As discussed in the report of the experiment \cite{Liedvogel2007}, transient absorption observed at 490-550~nm reveals a double exponential decay with decay times of ($4\pm1$)~ms and ($14\pm1$)~ms, whereas transient absorption at 550-630~nm shows only a single exponential decay with a decay time of ($14\pm1$)~ms. The decay times, therefore, may be attributed to $\mathrm{FADH}^{\bullet}$ ($14$~ms) and $\mathrm{FADH}^{-}$ (4~ms). The experimental error of the decay times is not provided in the original experimental papers and has been estimated here by digitizing the experimental data points and fitting the absorption curves with single and double exponential decays.

According to the original explanation of the recorded absorption patterns, \cite{Liedvogel2007} the double exponential decay of the transient absorption spectrum was attributed to the presence of the tryptophan radical in the sample for over 4~ms, and to the protein's signaling state associated with $\mathrm{FADH}^{\bullet}$ to be lasting over 14~ms. This explanation implies a radical pair lifetime in cryptochrome of over 1~ms, which is unusually long. However, a lifetime of the radical pair beyond 10~$\mu$s would abolish a magnetic field effect as singlet and triplet radical pair states $^1[\mathrm{FADH^{\bullet}+Trp^{\bullet}}]$ and $^3[\mathrm{FADH^{\bullet}+Trp^{\bullet}}]$, respectively, would then be dominated by electron spin relaxation that occurs faster than 10~$\mu$s~\cite{Rodgers2009}. For a significant magnetic field effect to arise, the radical pair lifetime needs to be short compared to electron spin relaxation times and, in fact, needs to be comparable with hyperfine interaction-induced singlet-triplet interconversion times of 1~$\mu$s~\cite{Rodgers2009}. At present there is no physical mechanism known that could explain a magnetic field effect for radical pair lifetimes $>10$~$\mu$s. The many suggestions for cryptochrome involvement in magnetoreception, for example the finding of Reppert \textit{et al.} for a cryptochrome knock-out mutant of fruit flies \cite{Gegear2008,Foley2011}, suggests to further pursue the possibility of a cryptochrome-based biochemical compass tied likely to a radical pair lifetime of $<10$~$\mu$s.

Such short lifetime seems to contradict the finding of Brettel \textit{et al.} \cite{GIOV2003} who isolated cryptochrome-1 from \textit{Arabidopsis thaliana} and studied the transient absorption of the protein; analyzing the change of the transient absorption the authors suggest that a tryptophan radical is present in cryptochrome for over 1~ms. However, this result cannot be applied to the present case since Brettel \textit{et al.} studied plant cryptochrome instead of avian cryptochrome. Although the photocycle of the members of the cryptochrome family should exhibit similarity, experiments show great variation between cryptochromes extracted from different organisms \cite{Bouly2007,Banerjee2007,Langenbacher2009,Berndt2007,Shirdel2008,Oeztuerk2008}. Furthermore, the change in the transient absorption spectra attributed by Brettel \textit{et al.} to the presence of the tryptophan radical may stem from another cryptochrome component with similar absorption properties. To explain an error due to such component we note that cryptochrome transient absorption relative to the absorbance prior to laser excitation can be calculated using Eq.~(\ref{eq:BeerLambertLaw}), where the summation is performed over the $\mathrm{FAD}$, $\mathrm{FADH}^{-}$ and $\mathrm{FADH}^{\bullet}$ states for a 490-550~nm probe-beam and includes only one term corresponding to the $\mathrm{FADH}^{\bullet}$ state in case of a 550-630~nm probe-beam. In the suggested model the transient $\mathrm{^1[FADH^{\bullet}+Trp^{\bullet}]}$, $\mathrm{^3[FADH^{\bullet}+Trp^{\bullet}]}$ states have no noticeable influence on the absorption spectrum since the lifetime of these radical pair states is assumed to be several orders of magnitude shorter than the lifetime of the $\mathrm{FADH}^{-}$ and $\mathrm{FADH}^{\bullet}$ states. If the lifetime of the radical pair state is on the order of milliseconds, as suggested in (\citenum{Liedvogel2007}), then the absorbance of the tryptophan radical becomes important.

Figure~\ref{fig:transientAbsorptionIntensity} shows the calculated time-dependence of the transient absorption in cryptochrome, obtained from solving rate equations (\ref{eq:rate_equations1})-(\ref{eq:rate_equations8}), for wavelengths at 490-550~nm and 550-630~nm. The figure shows that the energy of the laser pulse impacts the absorption properties of the protein. The dependence on the laser-pulse energy is due to the rate constant $k_{2}$ which is proportional to the laser pulse power (see Supplementary Material for more details). The transient absorption spectrum calculated for a pulse power of 1~MW, i.e., corresponding to the experimental 5~ns pulse carrying 5~mJ energy, can be compared to the experimental result reported earlier \cite{Liedvogel2007}. As can be seen in Fig.~\ref{fig:transientAbsorptionIntensity}, the calculated transient absorption perfectly matches the recorded data for a 490-550~nm and a 550-630~nm probe-beam.

Increasing the energy of the laser pump-pulse enhances the rate constant $k_{2}$ leading to faster population of the $\mathrm{FADH}^{-}$ state (see Fig.~\ref{fig:reaction}). This enhancement can be verified experimentally since the $\mathrm{FADH}^{-}$ and $\mathrm{FADH}^{\bullet}$ states in cryptochrome absorb light differently. Indeed, exciting the cryptochrome-containing sample with a pulse of higher intensity should lead to diminished absorption at 550-630~nm and a single exponential decay at 490-550~nm with decay time of 4~ms. Figure~\ref{fig:transientAbsorptionIntensity} illustrates how the time-dependence of the transient absorption behaves once the intensity of the pump pulse is increased.

A decrease of the pump-pulse power leads to suppression of the $\mathrm{FADH^{\bullet}}\rightarrow\mathrm{FADH^{\bullet*}}\rightarrow\mathrm{FADH^{-}}$ channel and extends the lifetime of the $[\mathrm{FADH^{\bullet}}+\mathrm{Trp^{\bullet}}]$ radical pair. Figure~\ref{fig:transientAbsorptionIntensity} shows the dependence of the transient absorption calculated for the laser pulse energy of 1~mJ acting over 5~ns. For such weak and short pulse the transition $\mathrm{FADH^{\bullet}}\rightarrow\mathrm{FADH^{\bullet*}}\rightarrow\mathrm{FADH^{-}}$ is significantly suppressed as the redistribution of the $\mathrm{FADH^{\bullet}}$ concentration towards the  $\mathrm{FADH^{-}}$ state is quenched. Figure~\ref{fig:transientAbsorptionIntensity} shows that for lower light intensities the absolute magnitude of the absorption at 550-630~nm increases, because the concentration of the $\mathrm{FADH}^{\bullet}$ states increases. Figure~\ref{fig:transientAbsorptionIntensity} presents experimental data only for laser pulse energy of 5~mJ; to establish unambiguity of the suggested model it is necessary to perform measurement at laser pulse energies of 1, 4, 7 and 10~mJ.

\begin{figure}[!t]
\begin{center}
\includegraphics[width=16cm]{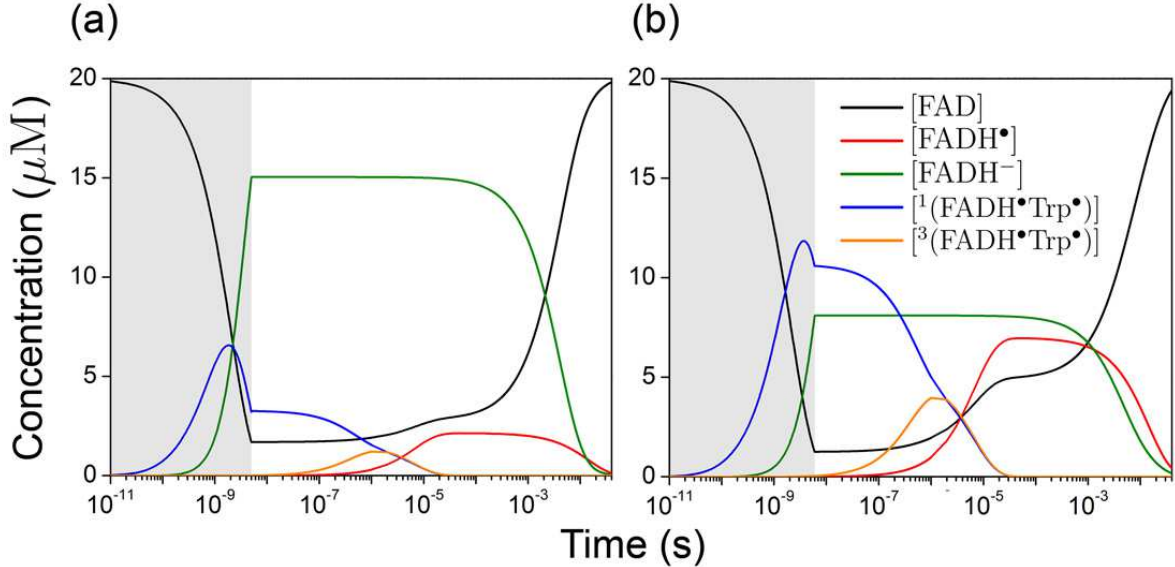}
\end{center}
\caption{
{\bf Population of transient states in cryptochrome.} Time evolution of concentration of the transient states in cryptochrome calculated from equations (\ref{eq:rate_equations1})-(\ref{eq:rate_equations8}) for cryptochrome photoexcitation by a laser pulse of 5~ns duration, 5~mJ  energy and 355~nm wavelength (plot a, consistent with Ref.~\citenum{Liedvogel2007}), and a laser pulse of 6~ns duration, 4~mJ energy and 460~nm wavelength (plot b, consistent with Ref.~\citenum{Biskup2009}). Grayed areas indicate time duration of laser pulse.}
\label{fig:ConcentrationEvolution}
\end{figure}

It is worth noting that the photoexcitation rate constants $k_1$ and $k_2$ depend not only on the laser-pulse energy, but also critically on the wavelength of the pulse and on pulse duration. Figure~\ref{fig:ConcentrationEvolution} presents the time evolution of the concentration of the transient states in cryptochrome calculated from Eqs.~(\ref{eq:rate_equations1})-(\ref{eq:rate_equations8}) for the photoexcitation regimes considered in one study \cite{Liedvogel2007} (plot a) and in another \cite{Biskup2009} (plot b). The difference in the pump-pulse leads to a significant change in the population of transient states. Thus, for a laser pulse of 5~ns duration, of 5~mJ energy, and 355~nm wavelength (see Fig.~\ref{fig:ConcentrationEvolution}a) the population of the $\mathrm{FADH^{-}}$ state after 5~ns is about 5 times higher than the population of the radical pair $^{1}[\mathrm{FADH^{\bullet}}+\mathrm{Trp^{\bullet}}]$ state. The early population of the $\mathrm{FADH^{-}}$ state readily explains both the double-exponential decay of the transient absorption spectrum at 490-550~nm and a single exponential decay at 550-630~nm as reported earlier \cite{Liedvogel2007}.

The absorption of the semiquinone state of the flavin radical ($\mathrm{FADH^{\bullet}}$) at 460~nm is significantly weaker in comparison to absorption at 355~nm (see Fig.~\ref{fig:flavinAbsorbtion} in Supplementary Material), and, therefore, the photoexcitation rate constant $k_2$ at 460~nm is smaller than at 355~nm. Figure~\ref{fig:ConcentrationEvolution}b shows that for a laser pulse of 6~ns duration, 4~mJ energy, and 460~nm wavelength the population of the $\mathrm{FADH^{-}}$ state after pulse excitation is smaller than the population of the radical pair $^{1}[\mathrm{FADH^{\bullet}}+\mathrm{Trp^{\bullet}}]$ state, arguing that the radical pair state can be better resolved at 460~nm than at 355~nm wavelength.

\subsection*{Influence of magnetic field as seen by transient spectra}

Comparison of measured and calculated kinetics in cryptochrome reveals good agreement and, hence, supports the reaction scheme in Fig.~\ref{fig:reaction}. Therefore, the postulated reaction scheme will be employed further to consider magnetic field effects on cryptochrome activation and deactivation.

The influence of an external magnetic field on cryptochrome photoactivation can be probed  through the effect of an applied magnetic field on the transient absorption spectrum. The magnetic field influences the singlet-triplet interconversion process in cryptochrome (see Fig.~\ref{fig:reaction}), described here extremely schematically through the first-order rate constant $k_{sf}$. An applied magnetic field leads to a relative change of the rate constant $k_{sf}$ \cite{SOLO2007,Ilia09,CINT2003} which brings about an altered transient absorption. The change can be written

\begin{equation}
\Delta A(\alpha,t)=A(k_{sf},t)-A(\alpha k_{sf},t),
\label{eq:DeltaA}
\end{equation}

\noindent
where $k_{sf}=10^{6}\ \mathrm{s^{-1}}$ is the reference value of the singlet-triplet interconversion rate constant (see Tab.~\ref{tab:rates}) and $\alpha$ denotes the relative change in $k_{sf}$ due to an altered magnetic field; $A$ is defined in Eq.~(\ref{eq:BeerLambertLaw}).

The $\alpha$-value depends on several factors, such as the hyperfine interaction in the radical pair, exchange and dipole-dipole interactions, and Zeeman interaction \cite{SOLO2007,Ilia09}. An accurate calculation of $\alpha$ requires the solution of a stochastic Liouville equation \cite{SOLO2007,Ilia09,CINT2003,Efimova2008}. We consider $\alpha$-values ranging from $\alpha=1.2$ ($k_{sf}=1.2\times10^6\ \mathrm{s^{-1}}$) to $\alpha=10$ ($k_{sf}=10^7\ \mathrm{s^{-1}}$), thereby accounting for realistic singlet-triplet interconversion rates expected in cryptochrome when the field reorients \cite{SOLO2007,Ilia09,Rodgers2009}.

Deviation of $\alpha$ from $\alpha=1$ can arise either through a change of magnetic field orientation or through changing field strength. The latter change can arise due to geographic variation of the geomagnetic field strength; this change is small percentage-wise, but is employed by birds apparently for a so-called magnetic map sense \cite{Semm90a,Alerstam87,Lednor88,MUNR1997}. In this case detection is likely achieved by an alternative receptor, an iron-mineral-based magnetoreceptor in the beak \cite{MUNR1997,Heyers2010,Kishkinev2010}. The geographic variation of the magnitude of the geomagnetic field is expected to have a negligible impact on the photo-reaction kinetics of cryptochrome and, therefore, is not considered here; only a change, i.e., $\alpha\ne1$, due to reorientation in the Earth' field is considered.

For cryptochrome to exhibit a response to the change of the magnetic field direction the protein needs be constrained orientation-wise within the organism. We have recently demonstrated that only one of three rotational degrees of freedom of cryptochrome need to be constrained to endow a bird with the magnetic compass \cite{SOLO2010}. Such constraint can be realized if cryptochromes are embedded in a cell membrane. The outer segment of the photoreceptor cells is an ideal structure to constrain the proteins as originally suggested in theory \cite{SOLO2010}. An experimental verification of cryptochromes localization in the outer segment of the UV photoreceptor cells was published recently \cite{Niessner2011}.

\begin{figure}[!t]
\begin{center}
\includegraphics[width=16cm]{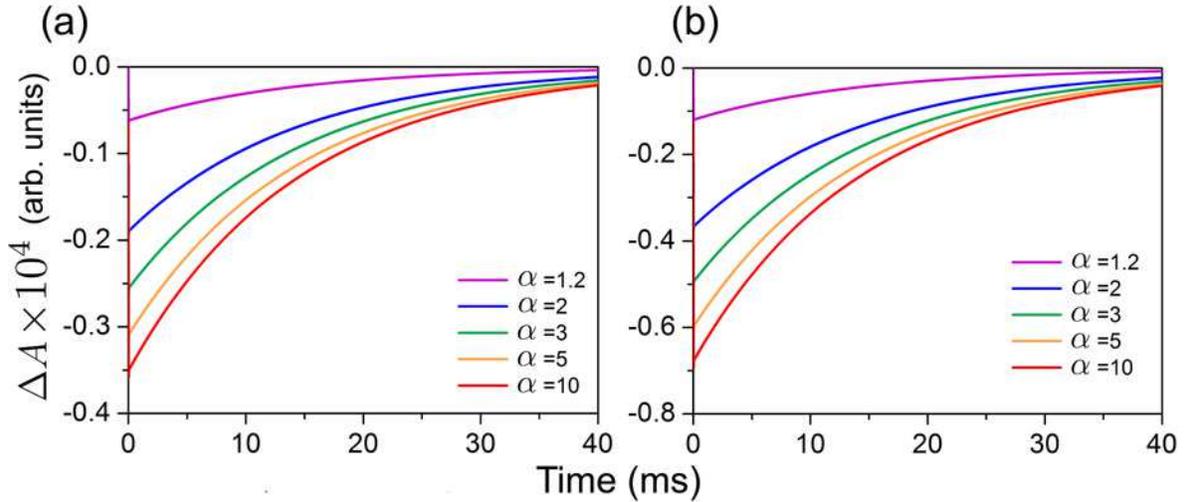}
\end{center}
\caption{
{\bf Magnetic field effect in cryptochrome.} Change of the transient absorption spectra in cryptochrome due to a change in $k_{sf}$ caused by an external magnetic field, calculated for the 550-630~nm probe-beam. $\alpha$ is defined in the text. Initial photoexcitation is due to a laser pulse of 5~ns duration, 355~nm wavelength and 5~mJ energy (plot a) and 2.5~mJ energy (plot b).  The curves in (a) and (b) reflect change in transient absorption due to an increased value of the $k_{sf}$ relative to the reference value $10^6$~s$^{-1}$ (see Tab.~\ref{tab:rates}).}
\label{fig:transientAbsorptionMagneticField}
\end{figure}

Figure~\ref{fig:transientAbsorptionMagneticField} shows the time-dependence of $\Delta A$ for various $\alpha$-values. The calculations were performed for cryptochrome photoexcitation by a laser pulse of 5~ns duration operating at 355~nm and having the energy of 5~mJ (see Fig.~\ref{fig:transientAbsorptionMagneticField}a) and 2.5~mJ for which the life-time of the $[\mathrm{FADH^{\bullet}}+\mathrm{Trp^{\bullet}}]$ radical pair is expected to be longer (see Fig.~\ref{fig:transientAbsorptionMagneticField}b). The relative change of $k_{sf}$ due to an applied magnetic field, $\alpha$, may vary within an order of magnitude as stated above and in Supplementary Material \cite{Rodgers2009,SOLO2007}.

Figure~\ref{fig:transientAbsorptionMagneticField} shows that decreasing the intensity of the pump-pulse leads to an enhancement of the magnetic field effect. The enhancement is due to the excitation channel $\mathrm{FADH^{\bullet}}\rightarrow\mathrm{FADH^{\bullet*}}\rightarrow\mathrm{FADH^{-}}$ in the cryptochrome photoactivation reaction (see Fig.~\ref{fig:reaction}) which becomes suppressed for the 2.5~mJ energy laser-pulse, such that the lifetime of the $[\mathrm{FADH^{\bullet}}+\mathrm{Trp^{\bullet}}]$ radical pair increases. Increasing  $k_{sf}$ $(\alpha>1)$ leads to faster singlet-triplet interconversion in the radical pair. If this interconversion becomes significantly faster than singlet and/or triplet decay, governed by rate constants $k_S$ and $k_T$ (see Fig.~\ref{fig:reaction}), the magnetic field effect in cryptochrome becomes negligible, as clearly seen in Fig.~\ref{fig:transientAbsorptionMagneticField}b, where the relative difference in $\Delta A$ calculated for $\alpha=5$ and $\alpha=10$ is small. A similar behavior is expected for higher pump-pulse intensities (see Fig.~\ref{fig:transientAbsorptionMagneticField}a) at higher values of $\alpha$, because in this case the $[\mathrm{FADH^{\bullet}}+\mathrm{Trp^{\bullet}}]$ radical pair decays predominantly through the $\mathrm{FADH^{\bullet}}\rightarrow\mathrm{FADH^{\bullet*}}\rightarrow\mathrm{FADH^{-}}$ channel, governed by the rate constant $k_{2}$, which is significantly larger than $k_S$ and $k_T$.

The dependencies shown in Fig.~\ref{fig:transientAbsorptionMagneticField} can also be measured experimentally and are important for resolving the physical mechanism of the magnetic field effect in cryptochrome. Similar measurements were performed earlier for DNA photolyase \cite{Henbest2008} and should be repeated for cryptochrome. The time-dependence of the transient absorption change due to a magnetic field allows one not only to judge whether the protein exhibits magnetoreceptive properties, but can also be used to reveal the impact of the magnetic field on the singlet-triplet interconversion rate constant $k_{sf}$.

Predicted absolute magnetic field effects in cryptochrome are rather small, i.e., the absorption changes only by about (0.1--0.3)$\times$10$^{-4}$ for a 5~mJ pulse (see Fig.~\ref{fig:transientAbsorptionMagneticField}a) and by about (0.2--0.6)$\times$10$^{-4}$ for a 2.5~mJ pulse (see Fig.~\ref{fig:transientAbsorptionMagneticField}b). Absolute changes of the transient absorption due to a 39~mT magnetic field in DNA photolyase were recorded to be of the order of $\sim$0.001 \cite{Henbest2008}, e.g., larger, but still very small. The relative magnetic field effect for DNA photolyase can thus be estimated to be $\sim$1~\%. The difference in the magnetic field effect recorded for DNA photolyase and predicted now for cryptochrome is due to the concentration of DNA photolyase used in experiment \cite{Henbest2008}, which was 0.2~mM, whereas in the present study the assumed concentration of cryptochrome is only 20~$\mu$M, following an earlier study \cite{Liedvogel2007}. Increasing the concentration leads to an enhancement of the absolute magnetic field effect.

In an animal the ambient light continuously irradiates cryptochromes in the retina and photo-excitation of $\mathrm{FADH^{\bullet}}$ induced through strong irradiation does not arise because the rate constant $k_2$ becomes small as compared to the value in the transient absorption experiment. \textit{In vivo} conditions lead then to an increase of the $[\mathrm{FADH^{\bullet}}+\mathrm{Trp^{\bullet}}]$ radical pair lifetime, increasing also the effect of an external magnetic field on cryptochrome signaling associated with $\mathrm{FADH^{\bullet}}$.

\begin{figure}[!t]
\begin{center}
\includegraphics[width=6in]{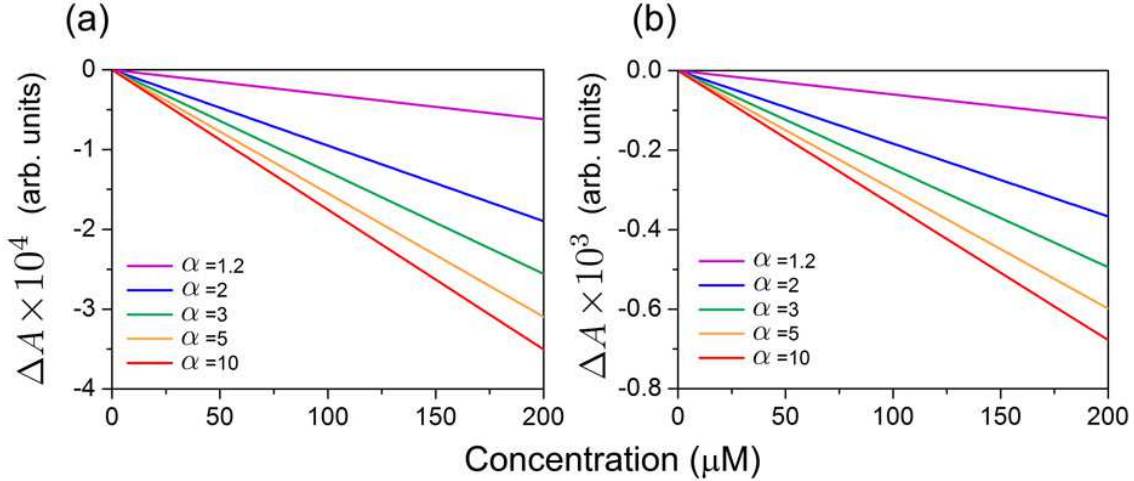}
\end{center}
\caption{
{\bf Magnetic field effect and probe concentration.} The maximal change of the transient absorption spectra in cryptochrome calculated due to change in $k_{sf}$ at different concentrations of protein, calculated for a 550-630~nm probe-beam. Initial photoexcitation  is due to a laser pulse of 5~ns duration, 355~nm wavelength, 5~mJ energy (plot a) or 2.5~mJ energy (plot b). The curves in (a) and (b) show transient absorption due to an increased  $k_{sf}$ relative to the reference value of $10^6$~s$^{-1}$ (see Tab.~\ref{tab:rates}), $\alpha$ is defined in the text.}
\label{fig:transientAbsorptionConcentration}
\end{figure}

Figure~\ref{fig:transientAbsorptionConcentration} shows the maximal change of the transient absorption due to a magnetic field effect calculated at different concentrations of the sample, and at different intensities of the pump-pulse (Fig.~\ref{fig:transientAbsorptionConcentration}a for 5~mJ over 5~ns and Fig.~\ref{fig:transientAbsorptionConcentration}b for 2.5~mJ over 5~ns). From the analysis in Fig.~\ref{fig:transientAbsorptionConcentration} follows that the magnitude of the maximal change of the transient absorption in cryptochrome increases linearly with the concentration of the sample. Thus, increasing the concentration to 200~$\mu$M could lead to a magnetic field effect of about 0.0007 (see Fig.~\ref{fig:transientAbsorptionConcentration}b), i.e. close to the value detected for DNA photolyase \cite{Henbest2008}. It is also worth noting that a more realistic account of the singlet-triplet interconversion predicts a relative magnetic field effect in cryptochrome of about 5-10~\% in a magnetic field comparable to the geomagnetic field \cite{SOLO2007,Ilia09}, which suggests that $\alpha$ is large, i.e. $\alpha\gg1$, and that the reference value of $k_{sf}$ is $\lesssim10^6\ \mathrm{s}^{-1}$.

The trivial concentration dependence of the maximal magnetic field effect in cryptochrome shown in Fig.~\ref{fig:transientAbsorptionConcentration} carries important information about the rate constants in cryptochrome photoactivation reaction. In particular, the slopes of the linear dependencies shown in Fig.~\ref{fig:transientAbsorptionConcentration} depend on these rate constants which govern the lifetime of transient states in the photoactivation reaction. The dependencies shown in Fig.~\ref{fig:transientAbsorptionConcentration} are convenient for the experiment, because the concentration of the cryptochrome-containing sample can be easily varied, and the transient absorption of the sample can then be recorded. Thus, by comparing results of the measurement with predictions of the kinetic model suggested in this paper it is feasible to justify rate constants in the photoactivation reaction, and predict the lifetime of certain intermediate states in cryptochrome.

The amount of cryptochrome in the bird's retina, presently unknown, is critical for the magnitude of the magnetic field effect. For a maximal signal a cryptochrome concentration larger than 20~$\mu$M is necessary. Figures~\ref{fig:transientAbsorptionMagneticField} and \ref{fig:transientAbsorptionConcentration} allow one to draw another important conclusion from our analysis, namely, that decreasing the intensity of the pump-pulse leads to an enhanced of the magnetic field effect. Interestingly, many migratory birds favor to fly at night time, i.e., when the intensity of the ambient light is low \cite{SOLO2010}.

\section{Conclusion}

Earlier investigations \cite{RITZ2000,Ilia09,SOLO2007,Ilia08b,SOLO2010} demonstrated how a biochemical magnetic compass in birds can be realized through light-induced electron transfer reactions in cryptochrome. However, the biophysical mechanism of cryptochrome activation and signaling still remains elusive. Our study provides an important step towards resolving the mechanism of cryptochrome magnetoreception.
%
%
The suggested analysis can provide a proof-of-principle whether cryptochrome is indeed the magnetoreceptor proteins in birds, comparing computed kinetic data with data from transient absorption spectroscopy, as successfully applied to DNA photolyase \cite{Henbest2008}.

The results of the present calculations reveal that magnetic field effects become only discernable if the concentration of the protein in the probe exceeds a critical value. Therefore, cryptochrome should be measured at different concentrations to ascertain that the minimal amount of protein is available. Transient absorption of cryptochromes at low concentration can be studied with the use of the so-called cavity-enhanced spectroscopy technique \cite{Banerjee2007}. This technique traps the probe light beam in an optical cavity, containing the sample, thereby transmitting the beam through the sample multiple times and increasing the transmittance length which in turn enhances the total absorption in the sample. The method permits protein concentrations below $\sim$1~$\mu$M \cite{Banerjee2007}, resolving nevertheless small changes in the transient absorption, as they arise due to an external magnetic field. A systematic study of the dependence of the absolute absorption of a cryptochrome-containing sample on cryptochrome concentration is important for revealing the lifetime of reaction intermediates through direct comparison of experimentally recorded absorption with the prediction of the kinetic formalism suggested in this paper.

The methodology suggested here illustrates how laser power, laser wavelength, laser pulse duration, probe light wavelength, and protein concentration, impact measured results. In particular, we identify which measurements pinpoint the magnetoreceptive properties of cryptochrome.

The performed study assumes for its analysis reaction rate constants, which were extracted from experiment or could be estimated otherwise. The uncertainty of most of the rate constants, is rather large. By measuring the transient absorption spectra in cryptochrome at different conditions, i.e., for different probe-beam intensities, different applied magnetic fields, different wavelengths of the pulse and probe beams, etc, and by comparing recorded profiles with calculated ones one can narrow the range of consistent rate constants in the proposed reaction scheme.

Magnetoreception in birds has been found to work for light with wavelengths of up to 565-567~nm \cite{Wiltschko1999,Muheim2002}. Thus, it is important to continue studies of cryptochrome photoactivation \textit{in vitro} by maintaining the conditions close to those used in the \textit{in vivo} experiments \cite{Wiltschko1999,Muheim2002}. An important question is how the magnetic directional information is disrupted under monochromatic light of an increased intensity. It has been demonstrated that different light intensities can disorient birds in the laboratory \cite{Wiltschko2000,WILT2001,Wiltschko2003,Muheim2002}, and, therefore, light intensity may affect the photocycle in cryptochrome. Studying transient absorption of cryptochrome in electromagnetic fields of certain frequencies, e.g., Larmor frequency of the electron \cite{RITZ2004,Hore2009,Ritz2009}, is another step to link \textit{in vitro} studies to behavioral investigations \cite{RITZ2004,Ritz2009}.

Science is still a long way from an explanation of avian magnetoreception. The best that may be said of our understanding today is that birds indeed perceive and use magnetic field information, and that their responses to magnetic fields under different conditions~--~light intensity and color, magnetic field strength and presence of oscillating fields~--~are suggesting a complex sensory system of multiple receptors. Even if cryptochrome is confirmed as a magnetoreceptor, it remains for biologists to determine how its magnetic field modulated signaling enters into a bird's sensory perception and ultimately into its orientation behavior.

\section{Supporting Material}
In Supporting Material we present details on flavin absorption spectra and discuss the choice of the rate constants in the studied reaction in cryptochrome.

\subsection{Absorption spectrum}

The efficiency of light absorption at wavelength $\lambda$ by an absorbing medium is characterized through the absorbance $A(\lambda)$ defined as \cite{Lakowicz2006}

\noindent
\begin{equation}
A(\lambda)=\log\frac{I_0(\lambda)}{I(\lambda)},
\label{eq:absorbance}
\end{equation}

\noindent
where $I_0(\lambda)$ and $I(\lambda)$ are the light intensities of the beam entering and leaving the absorbing medium, respectively. According to the reaction scheme in Fig.~\ref{fig:reaction}, cryptochrome can occupy several states, which are expected to absorb light differently. Therefore, according to the Beer-Lambert-Bouguer law \cite{Lakowicz2006}, the absorbance of the cryptochrome-containing sample is given by Eq.~(\ref{eq:BeerLambertLaw}).

The molar extinction coefficient, $\varepsilon_i$, in Eq.~(\ref{eq:BeerLambertLaw}) of a light-absorbing component in the system is directly related to the absorption cross section, $\sigma_i$, which characterizes the photon-capture area of a molecule \cite{Lakowicz2006}

\noindent
\begin{equation}
\sigma_i(\lambda) = 1000 \ln(10) \frac{\varepsilon_i(\lambda)}{N_A} = 3.82 \times 10^{-21} \varepsilon_i(\lambda).
\label{eq:absorption_cross_section}
\end{equation}

\noindent
Here $N_A$ is Avogadro's number; the absorption cross section in Eq.~(\ref{eq:absorption_cross_section}) is measured in units of $\mathrm{cm^{2}}$.

\subsection{Rate constants}

The rate constants in Eqs.~(\ref{eq:rate_equations1})-(\ref{eq:rate_equations8}) determine the time evolution of intermediate states in cryptochrome. Many of the rate constants are available from experiments performed on cryptochrome and cryptochrome-like proteins from various species, such as garden warbler \cite{Liedvogel2007}, \textit{Drosophila mela\-nogaster} \cite{Shirdel2008,Hoang2008}, \textit{Arabidopsis thaliana} \cite{Bouly2007,GIOV2003}, and \textit{Homo sapiens} \cite{Biskup2009,Hoang2008}. Many studies were also done for DNA photolyase, a protein structurally similar to cryptochrome \cite{Henbest2008,BYRD2003,Kao2008,AUBE2000,Gindt1999,AUBE2000,Prytkova2007}. Some of the rate constants can be independently estimated from fundamental physical principles \cite{SOLO2007,Ilia09}. The adopted rate constants are summarized  in Tab.~\ref{tab:rates}.

\noindent
\paragraph*{Flavin photoexcitation.}
\noindent
The rate constants $k_{1}$ and $k_{2}$ represent the rate of photoexcitation of the flavin cofactor from its fully oxidized FAD state and from the semiquinone $\mathrm{FADH^{\bullet}}$ state to the excited $\mathrm{FAD^{*}}$ and $\mathrm{FADH^{\bullet*}}$ states, respectively, (see Fig.~\ref{fig:reaction}). The $\mathrm{FAD}\rightarrow\mathrm{FAD^{*}}$ and $\mathrm{FADH^{\bullet}}\rightarrow\mathrm{FADH^{\bullet*}}$ transitions are induced by a laser pulse (see Fig.~\ref{fig:experiment}) and arise only during the pulse duration $\tau$. The rate constants $k_{1}$ and $k_{2}$ depend on the laser power and can be estimated as

\begin{equation}
k_{ex}=\sigma\frac{P\chi}{E_{ph}S_{0}}\left(1-\Theta(t-\tau)\right),
\label{eq:rateExcitation}
\end{equation}

\noindent
where $P=E/\tau$ is the power of the laser pulse (with $E$ being the energy of the pulse and $\tau$ the pulse duration), $S_{0}$ is the cross section area of the light beam hitting the sample, $\chi\le1$ defines the fraction of power deposited at the sample, $\sigma$ is the absorption cross section defined in Eq.~(\ref{eq:absorption_cross_section}), $E_{ph}=hc/\lambda$ is the energy of a single photon (with $h$ being the Planck constant and $c$ the speed of light), $\Theta(x)$ is the Heaviside step-function which limits the photoexcitation of FAD and $\mathrm{FADH^{\bullet}}$ to the period of the laser pulse duration. Here we do not consider the periodicity of the laser pulses as the time interval between two successive pulses is significantly longer than the typical reaction times involved in the scheme shown in Fig.~\ref{fig:reaction} (the pulse frequency used in the observation \cite{Liedvogel2007} was 10~Hz).

Assuming a Gaussian radial profile of the laser beam and substituting Eq.~(\ref{eq:absorption_cross_section}) into Eq.~(\ref{eq:rateExcitation}) one obtains

\begin{equation}
k_{ex}=612.686 \times\frac{P\lambda\varepsilon(\lambda)}{R^2}\left[\frac{1-\exp\left(-R^2/\omega^2\right)}{1-\exp\left(-R_{0}^2/\omega^2\right)}\right]\left(1-\Theta(t-\tau)\right),
\label{eq:rateExcitation2}
\end{equation}

\noindent
where $R$ is the radius of the beam at the sample measured in $\mu$m, $R_0$ is the radius of the output beam from the laser measured in $\mu$m and $\omega$ is the radius at which the laser field amplitude drops to $1/e$. $\varepsilon$ in Eq.~(\ref{eq:rateExcitation2}) is measured in $\mathrm{L\cdot mol^{-1}cm^{-1}}$, $\lambda$ is measured in $\mathrm{nm}$ and $P$ is measured in $\mathrm{Watt}$.

In the measurements \cite{Liedvogel2007} the $R_0$ value was $R_0=3000$~$\mu$m and the size of the sample used was likely smaller allowing one to assume $R=1500$~$\mu m$. With $\omega=1000$~$\mu$m, a typical value for the amplitude fall-off of the laser beam \cite{Bartels2006,Berera2009}, Eq.~(\ref{eq:rateExcitation2}) can be used to estimate the photoexcitation rate constants $k_1$ and $k_2$. The rate constants $k_{1}$ and $k_{2}$ are determined by the laser wavelength, $355\ \mathrm{nm}$, and the beam power, $10^{6}\ \mathrm{W}$ \cite{Liedvogel2007}. Figure~\ref{fig:flavinAbsorbtion} shows that the extinction coefficient of the oxidized flavin (FAD) at $\lambda=355\ \mathrm{nm}$ is about $7900\ \mathrm{L\cdot mol^{-1}cm^{-1}}$ \cite{Prytkova2007,Liedvogel2007}, resulting in $k_{1}=6.8\times10^8\ \mathrm{s^{-1}}$. The absorption (extinction) spectra in Fig.~\ref{fig:flavinAbsorbtion} were recorded for the three redox states of FAD \cite{SchmidtBook} and normalized to the absorption of FAD at 450~nm ($\varepsilon_{450}=11.3\times10^4\ \mathrm{L\cdot mol^{-1}cm^{-1}}$) (solid line) \cite{Liedvogel2007}. We note that the wavelength dependence of the extinction coefficient in garden warbler \cite{Liedvogel2007}, Drosophila, and human \cite{Biskup2009,Shirdel2008} cryptochromes maintains the general features of the absorption profile shown in Fig.~\ref{fig:flavinAbsorbtion}. Similarly, Fig.~\ref{fig:flavinAbsorbtion} shows that the extinction coefficient of semiquinone ($\mathrm{FADH^{\bullet}}$) flavin at $\lambda=355\ \mathrm{nm}$ is about $10400\ \mathrm{L\cdot mol^{-1}cm^{-1}}$ (solid line) \cite{Prytkova2007,Liedvogel2007}, resulting in $k_{2}=9\times10^8\ \mathrm{s^{-1}}$. The rate constants $k_1$ and $k_2$ apply only during the laser pulse duration time of $\tau=5$~ns.

\noindent
\paragraph*{Flavin excited states relaxation.}
\noindent
The rate constants $k_{rel}^{(1)}$ and $k_{rel}^{(2)}$ describe the $\mathrm{FAD^{*}}\rightarrow\mathrm{FAD}$ and $\mathrm{FADH^{\bullet*}}\rightarrow\mathrm{FADH^{\bullet}}$ relaxation processes, respectively (see Fig.~\ref{fig:reaction}). These relaxation processes have not been very well documented in cryptochrome, but the $\mathrm{FADH^{\bullet*}}\rightarrow\mathrm{FADH^{\bullet}}$ transition was studied in DNA photolyase \cite{BYRD2003}, a protein homologous to cryptochrome \cite{BRAU2004,LIN2003}. According to experimental measurement \cite{BYRD2003} the lifetime for the relaxation of $\mathrm{FADH^{\bullet*}}$ to the ground state is $80\ \mathrm{ps}$, leading to the value $k_{rel}^{(2)}=1.25\times10^{10}\ \mathrm{s^{-1}}$. The $\mathrm{FAD^{*}}\rightarrow\mathrm{FAD}$ transition is expected to occur on a similar timescale and, accordingly, we assume $k_{rel}^{(1)}=1.25\times10^{10}\ \mathrm{s^{-1}}$.

\noindent
\paragraph*{Radical pair formation.}
\noindent
The rate constant $k_{rp}$ describes the $^{1}[\mathrm{FADH^{\bullet}}+\mathrm{Trp^{\bullet}}]$ radical pair formation process (see Fig.~\ref{fig:reaction}). The characteristic time for this process is $30\ \mathrm{ps}$, as confirmed by using ultrafast pump-probe spectroscopy in the near-infrared spectral region in DNA photolyase \cite{AUBE2000}, leading to $k_{rp}=3.3\times10^{10}\ \mathrm{s^{-1}}$. Although $k_{rp}$ has not been clearly resolved for cryptochrome, we assume the rate constant from DNA photolyase to be of the same order of magnitude in all photolyase/cryptochrome-like proteins \cite{Henbest2008,Biskup2009,SOLO2007}.

\noindent
\paragraph*{Flavin deprotonation.}
\noindent
The rate constant $k_{ox}^{(1)}$ describes the deprotonation process of $\mathrm{FADH^{+}}$, as denoted in Fig~\ref{fig:reaction}. Since the $\mathrm{FADH^{+}}$ state of the flavin cofactor has never been observed in cryptochrome, and/or photolyase \cite{AUBE2000,Henbest2008,GIOV2003,Bouly2007,BYRD2003,AUBE2000,Kao2008,Ritz2010,Hoang2008,Liedvogel2007,Gindt1999,Biskup2009}, the characteristic time of $\mathrm{FADH^{+}}$ deprotonation is expected to be on the order of few picoseconds, which is beyond experimental resolution. Thus, we assume the value $k_{ox}^{(1)}=1/10\ \mathrm{ps}=10^{11}\ \mathrm{s^{-1}}$.

\noindent
\paragraph*{Electron transfer involving flavin radical.}
\noindent
The rate constants $k_{et}^{(1)}$ describes electron transfer from the $\mathrm{FADH^{\bullet}}$ radical to the $\mathrm{Trp^{\bullet}}$ radical (see Fig.~\ref{fig:reaction}). $k_{et}^{(1)}$ is expected to have an approximate value of $1/10\ \mathrm{\mu s}=10^{5}\ \mathrm{s^{-1}}$, since the radical pair in cryptochrome is assumed to have a lifetime of $6\ \mathrm{\mu s}$ \cite{Biskup2009}; measurements were performed using transient EPR spectroscopy, with the system optically excited by a Nd:YAG laser (Spectra Physics GCR-11) pumping an optical parametric oscillator (Opta BBO-355-vis/IR, Opta GmbH, Bensheim, Germany) tuned to a wavelength of 460 nm (pulse width 6 ns; pulse energy 4 mJ). At this particular wavelength the extinction coefficient of the semiquinone ($\mathrm{FADH^{\bullet}}$) flavin is about $3600\ \mathrm{L\cdot mol^{-1}cm^{-1}}$ (see blue solid line in Fig.~\ref{fig:flavinAbsorbtion}). The chosen laser pulse power leads to the photoexcitation rate constant $k_{2}=2.6\times10^8\ \mathrm{s^{-1}}$ under the assumption that the geometrical characteristics of the laser beams in two experiments \cite{Liedvogel2007,Biskup2009} are identical. Since the fast photoexcitation decay channel of the radical pair is only possible during the period of the laser pulse, the fourfold decrease of the photoexcitation rate constant would lead to a significant increase of the lifetime of the radical pair state allowing its detection in the EPR measurements.

The rate constant $k_{et}^{(2)}$ describes electron transfer from the $\mathrm{Trp^{\bullet}}$ radical to the excited $\mathrm{FADH^{\bullet*}}$ radical. The electron transfer rate constant for this process was measured in DNA photolyase using time-resolved absorption spectroscopy and found to be $k_{et}^{(2)}=2.6\times10^{10}\ \mathrm{s^{-1}}$ \cite{BYRD2003}. Similar values for the electron transfer rate constants can also be estimated from Marcus theory of electron transfer \cite{SOLO2007}.

\noindent
\paragraph*{$\mathrm{S}\leftrightarrow\mathrm{T}$ interconversion.}
\noindent
The rate constant $k_{sf}$ describes the singlet-triplet interconversion process in cryptochrome (see Fig.~\ref{fig:reaction}) as a first order reaction process. We employ such process as a very rough model for the actual quantum mechanical spin precession process, since only time scale and yield of singlet-triplet interconversion matter. The $k_{sf}$ rate constant depends on several factors, such as the hyperfine interaction in the radical pair, exchange and dipole-dipole interaction between the radical pair partners, and the external magnetic field. To calculate the singlet-triplet kinetics one needs to solve the stochastic Liouville equation for the radical pair in the system \cite{SOLO2007,Ilia09,CINT2003,Efimova2008}. In the present study we cast the transition process into a single rate constant as done earlier \cite{SCHU76C}. It has been demonstrated \cite{SOLO2007,Ilia09,Rodgers2009} that for a generic radical pair holds $k_{sf}=10^{6}-10^{8}\ \mathrm{s^{-1}}$. $k_{sf}$ is magnetic field dependent and, thereby, responsible for the magnetic field effect in cryptochrome. We demonstrate the feasibility of a magnetic field effect in cryptochrome by varying the value of $k_{sf}$. If cryptochrome is the primary magnetoreceptor protein in birds and other animals, it is natural to assume that nature has designed it in such a way that the external geomagnetic field produces a significant effect. The time needed for a significant transformation of a singlet state $^{1}[\mathrm{FADH^{\bullet}}+\mathrm{Trp^{\bullet}}]$ into a triplet state $^{3}[\mathrm{FADH^{\bullet}}+\mathrm{Trp^{\bullet}}]$ and vice versa in a 0.5~G magnetic field is typically $\sim700\ \mathrm{ns}$ \cite{Rodgers2009}; therefore we assume $k_{sf}=10^{6}\ \mathrm{s^{-1}}$ for the singlet-triplet interconversion rate constant.

\noindent
\paragraph*{Singlet and Triplet decay kinetics.}
\noindent
The rate constants $k_{S}$ and $k_{T}$ describe the singlet and the triplet decays of the $[\mathrm{FADH^{\bullet}}+\mathrm{Trp^{\bullet}}]$ radical pair (see Fig.~\ref{fig:reaction}). Together with the $k_{et}^{(1)}$ electron transfer rate constant they define the lifetime of the radical pair. According to the experiment, the lifetime of the radical pair in cryptochrome is $\gtrsim6\ \mathrm{\mu s}$ \cite{Biskup2009}; therefore, we assume $k_{S}=k_{T}=1/10\ \mathrm{\mu s}=10^{5}\ \mathrm{s^{-1}}$, i.e. we take the lower bound as our estimate of the radical pair lifetime. For the sake of simplicity we also assume spin-independent decay kinetics of the radical pair, i.e. $k_{S}=k_{T}$.

\noindent
\paragraph*{Dark reaction kinetics.}
\noindent
The rate constants $k_{red}$ and $k_{ox}^{(2)}$ are associated with the two-step cryptochrome dark reaction (see Fig.~\ref{fig:reaction}). Both stages involved are expected to be fairly long-lived, with a lifetime on the order of milliseconds, and should be resolved in transient absorption spectra reported in the experiment \cite{Liedvogel2007}. According to the experiment, two transient states in cryptochrome with lifetimes of $4\ \mathrm{ms}$ and $14\ \mathrm{ms}$ were detected. $k_{red}$ describes the reduction process of the semiquinone $\mathrm{FADH^{\bullet}}$ radical possibly involving the $\mathrm{O}_2^{\bullet-}$ radical, as suggested earlier \cite{Ilia09}. It is natural to expect that the lifetime of the signalling state in cryptochrome is maximal, leading to the values $k_{red}=1/14\ \mathrm{ms}\approx70\ \mathrm{s^{-1}}$ and $k_{ox}^{(2)}=1/4\ \mathrm{ms}=250\ \mathrm{s^{-1}}$.

\subsection{Flavin absorption spectra}

\begin{figure}[!ht]
\begin{center}
\includegraphics[width=16cm]{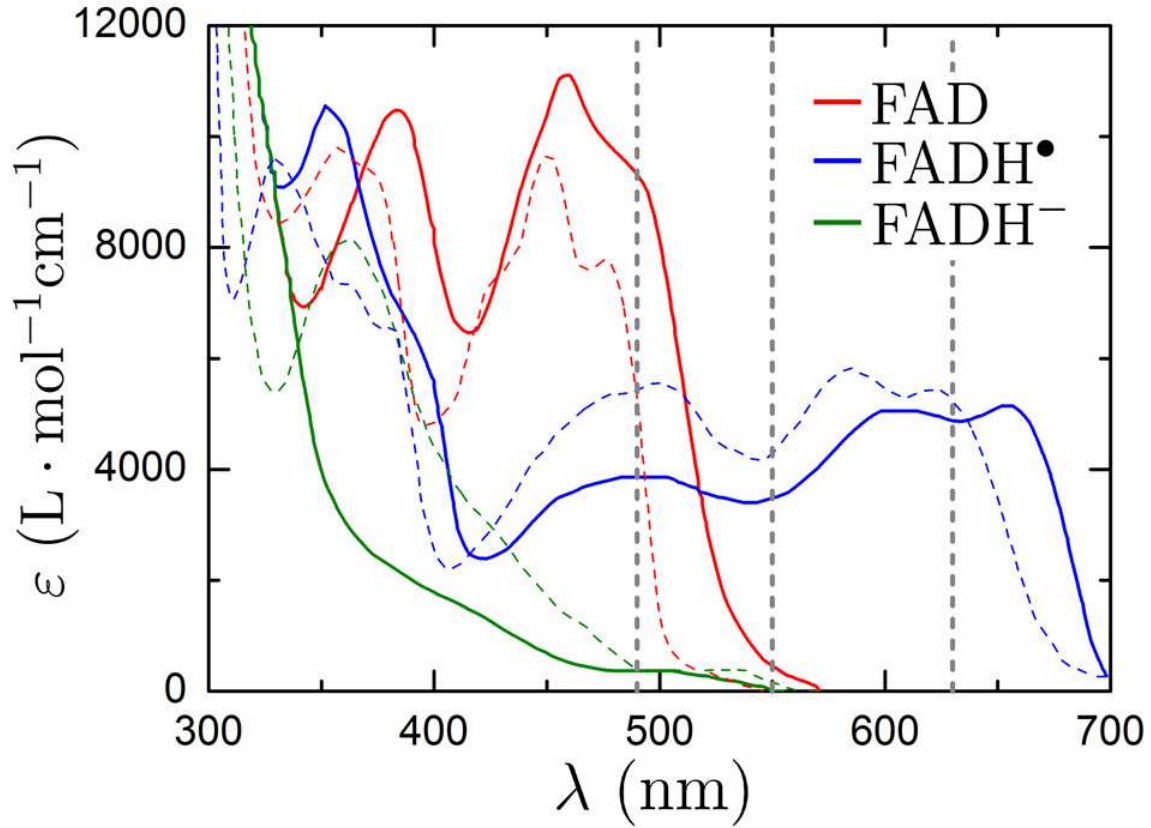}
\end{center}
\caption{
{\bf Flavin absorption spectra.} Wavelength dependence of the molar absorption of the flavin chromophore in its three redox states $\mathrm{FAD}$ (red line), $\mathrm{FADH}^{\bullet}$ (blue line), and $\mathrm{FADH}^{-}$ (green line). Vertical dashed lines set the limits for the wavelengths used in experiment \cite{Liedvogel2007} and in the calculations reported here. The spectra of the flavin chromophore shown as a solid line have been adopted from a textbook \cite{SchmidtBook}, while the spectra of different redox forms of selected cryptochrome/photolyase shown as a dashed line are digitized from the measured pattern \cite{Liu2010}.}
\label{fig:flavinAbsorbtion}
\end{figure}

Optical absorbance of cryptochrome is dominated by the absorbance of the isoalloxazine moiety in FAD \cite{Liedvogel2007,Biskup2009}. Figure~\ref{fig:flavinAbsorbtion} shows absorption spectra recorded for the three redox states of FAD: the fully reduced $\mathrm{FADH}^{-}$, one-electron oxidized $\mathrm{FADH}^{\bullet}$ and the two-electron oxidized $\mathrm{FAD}$ \cite{Biskup2009,SchmidtBook,Liu2010}.

The calculated absorption spectrum in Fig.~\ref{fig:transientAbsorptionIntensity} depends on the value of the extinction coefficients, $\varepsilon_i$, of the cryptochrome transient states (see Eq.~(\ref{eq:BeerLambertLaw})), which are wavelength dependent. The extinction coefficients for $\mathrm{FAD}$, $\mathrm{FADH}^{\bullet}$ and $\mathrm{FADH}^{-}$ were chosen as $300\ \mathrm{L\cdot mol^{-1}cm^{-1}}$, $2695\ \mathrm{L\cdot mol^{-1}cm^{-1}}$ and $690\ \mathrm{L\cdot mol^{-1}cm^{-1}}$, respectively, for 490-550~nm, and the extinction coefficient for $\mathrm{FADH}^{\bullet}$ at 550-630~nm was assumed to be $3843\ \mathrm{L\cdot mol^{-1}cm^{-1}}$. Figure~\ref{fig:flavinAbsorbtion} illustrates that the chosen values are consistent with the experimentally recorded absorption profiles for the different redox states of the flavin moiety.

The only noticeable difference between calculated and observed spectra is the increased value of the extinction coefficient for the fully reduced $\mathrm{FADH}^{-}$ state of the flavin cofactor, which in the calculation is slightly larger than the extinction coefficient for the fully oxidized $\mathrm{FAD}$ state. However, both extinction coefficients are expected to be small (see Fig.~\ref{fig:flavinAbsorbtion}), as they are taken from the far edge of the absorption spectrum. Therefore, the inaccuracy in their value is high, but inconsequential. To illustrate the uncertainty of flavin extinction at 490-550~nm in Fig.~\ref{fig:flavinAbsorbtion} we show two sets of absorption profiles for the different redox states of the flavin cofactor recorded for an isolated flavin chromophore \cite{SchmidtBook} (solid line) and for the selected cryptochrome/photolyase proteins \cite{Liu2010} (dashed line). Although in both systems the absorption spectra contain similar features, one notes that the spectra for the oxidized (FAD) and semiquinone ($\mathrm{FADH}^{\bullet}$) states in the isolated flavin chromophore are red-shifted with respect to the spectra recorded for cryptochrome/photolyase proteins, thereby exhibiting significantly different absorption at 490-550~nm. Another reason for the increased absorption of the $\mathrm{FADH}^{\bullet}$ state in our model may be that in the experiment \cite{Liedvogel2007} $\mathrm{FADH}^{-}$ is actually substituted by the anionic $\mathrm{FAD}^{\bullet-}$ radical state which exhibits a somewhat higher absorption at 490-550~nm \cite{Liu2010}. This explanation is worth studying in a systematic fashion.

Transient absorption is a powerful technique for studying intermediate states in chemical reactions, but the method often delivers data which cannot be unequivocally interpreted since the absorption patterns for different molecules and their various states often overlap. The impact of simultaneous absorption of several transient states at the same wavelength on the total absorption of the sample had to be addressed already three decades ago for pyrene/DMDMA complex \cite{SCHU76C}, where a theoretical approach, similar to the method which we now use for cryptochrome transient absorption calculation, was successfully employed in the first quantitative demonstration of a radical pair magnetic field effect. Large biomolecules usually have many constituents which respond to the typical wavelengths used by probe light in the transient absorption experiment. Recent studies on transient absorption of cryptochrome, and cryptochrome-like proteins, often deliver a single explanation for the measured data and do not discuss how the result would change if other possible transient states of the studied proteins were taken into account. The only way to unequivocally interpret the results of measurements in the present case is through measurements at different conditions that likely impact the reaction kinetics and shift the equilibrium toward different transient states.

\begin{acknowledgments}
The authors thank Paul Galland (University of Marburg) for providing the necessary data on flavin absorption, Robert Bittl (Freie Universit{\"a}t Berlin) for stimulating discussions which helped us understand the experimental aspects of the transient absorption spectroscopy, Peter Hore and colleagues (University of Oxford) for providing the necessary information on the experimental setup used in cryptochrome transient absorption experiments and Vita Solovyeva for inspiring remarks. This work has been supported by National Science Foundation grants NSF MCB-0744057 and NSF PHY0822613 and by National Institutes of Health grant P41-RR005969. K.S. thanks the Alexander von Humboldt Foundation for support and I.S. acknowledges support as a Beckman Fellow.
\end{acknowledgments}

\bibliography{D:/Ilia/Papers/Bibliography/journals_short,D:/Ilia/Papers/Bibliography/Birds_04.05.11}

\end{document}